\documentclass[conference, final]{IEEEtran}

\usepackage{cite}
\usepackage{amsmath}
\usepackage{amsfonts}
\usepackage{amssymb}
\usepackage{makeidx}
\usepackage{mathrsfs}
\usepackage{amsthm}
\usepackage{url}
\usepackage{color}

\ifCLASSINFOpdf
   \usepackage[pdftex]{graphicx}
\else
\fi

\newtheorem{theorem}{Theorem}[]

\allowdisplaybreaks
\begin{document}
\title{Cellular Systems with Many Antennas: Large System Analysis under Pilot Contamination}

\author{    \authorblockN{Narayanan Krishnan, Roy D. Yates, Narayan B. Mandayam}
\authorblockA{WINLAB; Rutgers, The State University of New Jersey\\
E-mail: narayank, ryates, narayan @winlab.rutgers.edu}
    }

\maketitle \thispagestyle{empty}
\begin{abstract}
Base stations with a large number of transmit antennas have the potential to serve a large number of users simultaneously at higher rates. They also promise a lower power consumption due to coherent combining at the receiver. However, the receiver processing in the uplink relies on the channel estimates which are known to suffer from pilot interference. In this work, we perform an uplink large system analysis of multi-cell multi-antenna system when the receiver employs a matched filtering with a pilot contaminated estimate. We find the asymptotic Signal to Interference plus Noise Ratio (SINR) as the number of antennas and number of users per base station grow large while maintaining a fixed ratio. To do this, we make use of the similarity of the uplink received signal in a multi-antenna system to the representation of the received signal in CDMA systems. The asymptotic SINR expression explicitly captures the effect of pilot contamination and that of interference averaging. This also explains the  SINR performance of receiver processing schemes at different regimes such as instances when the number of antennas are comparable to number of users as well as when antennas exceed greatly the number of users. Finally, we also propose that the adaptive MMSE symbol detection scheme, which does not require the explicit channel knowledge, can be employed for cellular systems with large number of antennas.
\end{abstract}

\normalsize

\IEEEpeerreviewmaketitle
\section{Introduction}
Cellular systems with large number of base station antennas have been found to be advantageous in mitigating the fading effects of the channel \cite{Marzetta2011}. In the downlink, dense low-powered base stations operating with power in the order of milliwatts have the potential to conserve power as compared to the present systems. In the uplink, coherent receiver processing with a large number of antennas reduces the transmitted powers of the users. It is shown in \cite{Marzetta2011} that in an infinite antenna regime, and in a bandwidth of $20$~MHz,  a time division duplexing system has the potential to serve $40$ single antenna users with an average throughput of $17$~Mbps per user. 

However, any advantages offered by multiple transmitters at the base station can be utilized only by gaining the channel knowledge between the base station and all the users. This requires training data to be sent from the users. Hence, a part of the channel coherence time is utilized for gaining the channel knowledge between the base station and all users. In a typical system the time-frequency resources are divided into blocks of coherence-time coherence-bandwidth product, where some resources(time or frequency) are used for channel estimation and the rest is used for transmission in uplink or downlink. However, in \cite{Marzetta2006}, it is shown that the number of pilot symbols required is proportional to the total number of users in the system. Hence, as the system scales with the number of users, the dedicated training symbols may take up the coherence time of the channel.  As this is undesirable, only a part of the coherence time is utilized to learn the channel. As a result, the pilot sequences in different cells overlap over time-frequency resource and as a consequence the channel estimate is corrupted. This is called pilot interference which is found to be a limiting factor as we increase the number of antennas \cite{Jubin}. 

It is shown in \cite{Marzetta2011} that in the limit of infinite number of antennas, the SINR using a matched filter receiver is limited by interference power due to pilot contamination. While the result assumes a regime with finite number of users, we can also envision a regime where the number of users may be comparable to the number of antennas such as a system with $50$ antenna base stations serving $50$ users simultaneously. In this work, we do a large system analysis of uplink multi-cell, multi-antenna system when the receiver employs a matched filter to decode the received signal. We let the number of antennas and the number of users per base station grow large simultaneously while maintaining a fixed users to antennas ratio and observe the SINR in the following cases, 
\begin{itemize}
\item[1.] when there is a perfect channel estimate,
\item[2.] when we have a pilot corrupted channel estimate. 
\end{itemize}
In order to accomplish that we make use of the similarity of the uplink received signal in a MIMO system to that of the received signal in a CDMA system \cite{TseHanly1999}. Further we compare the results of the asymptotic SINR expression so obtained.

Also, we propose an adaptive filtering method reminiscent of CDMA systems where the uplink receiver filter at the base station converges to the desired MMSE filter. While MMSE filtering requires us to estimate the channel to all the users in the system, the adaptive MMSE does not require any prior estimation of the channel to any of the users. Using independent training symbols among users, the receive filter will converge to the desired MMSE filter.  

\section{System Model}
\label{sec:SystemModel}
We consider a system similar to that in \cite{Marzetta2011} with $B$ base stations and $K$ users per base station. We assume that all the $KB$ users in the system are allocated the same time-frequency resource. Also, each base station is equipped with $M$ antennas. The channel vector representing the small scale fading between  user $k$ in cell $j$ and the antennas in base station $l$ is given by a $M \times 1$ vector $\mathbf{h}^{(l)}_{jk}$. The entries of $\mathbf{h}^{(l)}_{jk}$ are assumed to be independent zero mean i.i.d Gaussian random variables with unit variance. This corresponds to an ideal and favourable propagation medium with rich scattering. The large scale fading coefficients, which represents the power attenuation due to distance and effects of shadowing between base station $l$ and $k^{th}$ user in $j^{th}$ cell is given by $\beta^{(l)}_{jk}$. This is constant across the antennas of the cell $l$. Accordingly, overall channel vector is given by $\mathbf{g}^{(l)}_{jk} = \sqrt{\beta^{(l)}_{jk}}\mathbf{h}^{(l)}_{jk} $. 

\subsection{Uplink Transmission}
 We assume that all user's transmission are perfectly synchronized.  Also, while a user's transmission is intended to its base station, other base stations also hear the transmission. Defining $q_{jk}$ as the symbol transmitted by user $k$ in cell $j$,
$\mathbf{w}^{(l)}$  as the $M\times 1$ noise vector with zero mean unit variance gaussian entries, $\rho_r/M$ as the uplink signal to noise ratio scaled by the number of antennas, the received signal at base station $l$ is given by,
\begin{eqnarray}
\mathbf{y}^{(l)} &=& \sqrt{\frac{\rho_r}{M}} \sum^B_{j=1} \sum^K_{k=1}\mathbf{g}^{(l)}_{jk}q_{jk}   + \mathbf{w}^{(l)} \label{eq:RcvdSgnlVectorForm}\text{.} 
\end{eqnarray}
 Since processing power is not an issue at the base station, received signal processing can be employed. However, the base station has to have an estimate of the channel to all users prior to transmission of uplink information.  In a system employing OFDM physical layer with time-frequency resources, we can divide the resources into blocks spanning the coherence-time coherence-bandwidth product. Although the channel vector $\mathbf{h}^{(l)}_{jk}$ to each of the users has to be relearned by the base station at the start of every coherence time, once learnt for a subcarrier it remains the same for of all subcarriers spanning the coherence bandwidth during that coherence time. Let the number of coherent symbols be given by $T_c$ and the coherent subcarriers be $N_c$. Therefore, if we fix the number of symbols used for estimation to be $T$ such that $T \leq T_c$, a total of $N_cT$ user's channel can be learnt. This observation was noted in \cite{Marzetta2011}. We would like to point out that it is relevant here as the number of users than can be supported depends on $N_c$ and depending on its value the number of users $K$ that could be supported can be comparable to $M$. Therefore, it is worthwhile to investigate not only $M \gg K$ scenario but also the case when $M$ and $K$ large and comparable. 

\subsection{Limitations in gaining Channel Knowledge }
During each coherence time, users in a cell spend some pilot symbol times over the subcarriers spanning the coherence bandwidth for channel estimation at the base station and then the transmission of data ensues until the end of the coherence time. At base station $l$, the number of channel vectors $\mathbf{h}^{(l)}_{jk}$  that needs to be learnt is equal to the number of users in the system which is $KB$ where $K = N_cT$. In order to accomplish that, the number of pilots required must at least be $KB$ symbol times. However, such a system will not be scalable as there exists some large $B$ for which the product $KB$ will occupy all the coherence time. This is clearly  undesirable as pilot training is taking up significant part of a coherence time.   

\subsection{Pilot Interference based Channel Estimate}
In one of the approaches taken in \cite{Marzetta2011}, the base station is concerned with only knowing the channel to its own $K$ users and spends only $K$ time-frequency resources for channel estimation instead of $KB$. Every base station similarly spends first $K$ time-frequency resources for channel estimation to its $K$ users. Intuitively, we can think of  $K$ users of the $l^{th}$ base transmitting pilots in $K$ orthogonal times-frequency resource to its base. Consequently, the user indexed one will be transmitting pilot symbol to its base station $l$ in the first time slot in a subcarrier. Since the first user in other cells also transmits at the same time and at the same subcarrier, the received signal is corrupted by other pilots transmission.

This results in an estimate as given below, 
\begin{eqnarray}
\hat{\mathbf{h}}^{(l)}_{lk} = \mathbf{g}^{(l)}_{lk} +  \sqrt{\kappa} \sum_{j \neq l} \mathbf{g}^{(l)}_{jk} \label{eq:Channelest},
\end{eqnarray}
The variable $\kappa$ which can take values $0$ or $1$ is in order to distinguish a perfect estimate and a pilot corrupted estimate respectively. Since our aim is to analyse the effect of pilot interference we assume that there  is no uncorrelated gaussian noise further corrupting the estimate. It has been shown in \cite{Marzetta2011,Jubin} that using this estimate and a matched filter decoder the achievable rates are limited by inter-pilot interference as the number of antenna grows.

\section{Large System Analysis}
Since the SINR analysis is identical to all users in the system we focus only on user $1$ in base station indexed $1$. For matched filtering we project the received signal $\mathbf{y}^{(1)}$ in equation (\ref{eq:RcvdSgnlVectorForm}) onto the normalized channel estimate $\hat{\mathbf{g}}^{1}_{11}/\sqrt{M}$. We also drop the superscript $(.)^{(1)}$ for notational convenience. Consequently, the SINR of the first user to its base station with pilot corrupted channel estimate is given by,
\begin{eqnarray}
\label{eq:SINRMF}
\mathsf{SINR}&=& \frac{  \rho_r \left|\frac{\hat{\mathbf{g}}^{H}_{11} \mathbf{g}_{11} }{M} \right|^2      }{\frac{\hat{\mathbf{g}}^{H}_{11} \hat{\mathbf{g}}_{11}}{ M} +  \rho_r \underset{\{(j,k) \neq (1,1)\}}{\sum^B_{j=1} \sum^K_{k=1}} \left|\frac{\hat{\mathbf{g}}^{H}_{11} \mathbf{g}^{}_{jk} }{M} \right|^2 }
\end{eqnarray}

%
\begin{theorem} 
\label{th:SINRLargeSystem}
Let $\mathsf{SINR}$ obtained with matched filtering be as given in equation (\ref{eq:SINRMF}), then as $M,K \rightarrow \infty$, with $K/M = \alpha$, $\mathsf{SINR}$ converges in probability to
\begin{eqnarray}
\label{eq:SINRLargeSystem}
\mathsf{SINR}^{*} = \frac{   \rho_r \frac{ \beta_{11} }{ 1 +  \kappa \left( \sum^{B}_{j=2}\beta_{j1} \right)/\beta_{11} }   }{ 1 +   \rho_r    \left[ \kappa \frac{ \left(\sum^{B}_{j=2}\beta^2_{j1} \right)/\beta_{11}}{1 + \kappa \left(\sum^{B}_{j=2} \beta_{j1} \right)/\beta_{11} } + B\alpha \mathbb{E}[\beta]\right]  }
\end{eqnarray}
where, the expectation is over the distribution of $\beta$ - the random variable representing the realizations of large scale fading gains $\beta_{jk}'s$.
\end{theorem} 
\begin{IEEEproof}
Proof given in Appendix
\end{IEEEproof}

The expression for the asymptotic SINR captures the  effect of both interference due to pilot contamination and the interference averaging due to $\alpha \neq 0$. The SINR expression in the limit of infinite antennas but finite number of users per cell are obtained when we put $\alpha =0$ and this corresponds to the expression for SINR in \cite{Marzetta2011, Jubin}. It is seen that the SINR expression so obtained is limited by the pilot interference powers. On the other hand with a perfect channel estimate,
$\hat{\mathbf{g}}^{(l)}_{lk} = \mathbf{g}^{(l)}_{lk}$ implying that $\kappa = 0$ we get the SINR of matched filter as in \cite{TseHanly1999}. We can view Theorem \ref{th:SINRLargeSystem} as a generalization of the large system analysis for a  matched filter receiver with a pilot contaminated channel estimate. 

\section{Numerical Evaluation}
\begin{figure}[h]
    \begin{center}
      \includegraphics[width = 3in, clip = true, trim = 100 240 100 240]{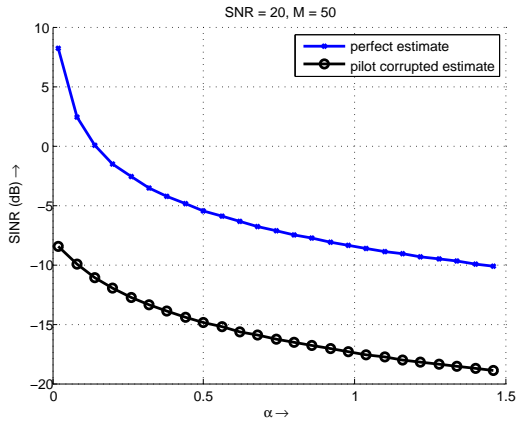}
      \caption{ The asymptotic limit of matched filter SINR  with a perfect estimate is plotted with that of matched filter with a pilot corrupted estimate. The number of antennas considered is $50$ and the signal to noise ratio is $20$~dB. } 
      \label{fig:AsyPlots}
    \end{center}
\end{figure}

\begin{figure}[h]
    \begin{center}
      \includegraphics[width = 3in, clip = true, trim = 100 240 100 240]{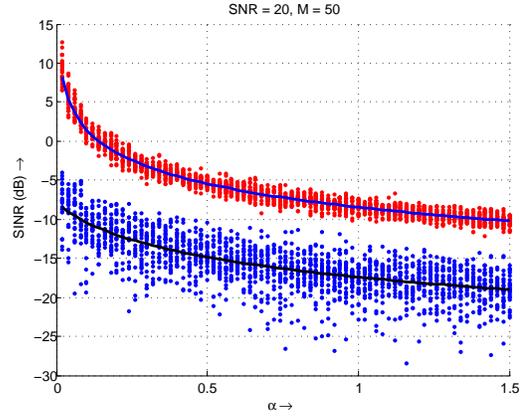}
      \caption{ We observe that the random realizations of the SINRs are clustered around the asymptotic limit. } 
      \label{fig:m50_TH}
    \end{center}
\end{figure}

\begin{figure}[h]
    \begin{center}
      \includegraphics[width = 3in, clip = true, trim = 10 0 10 0]{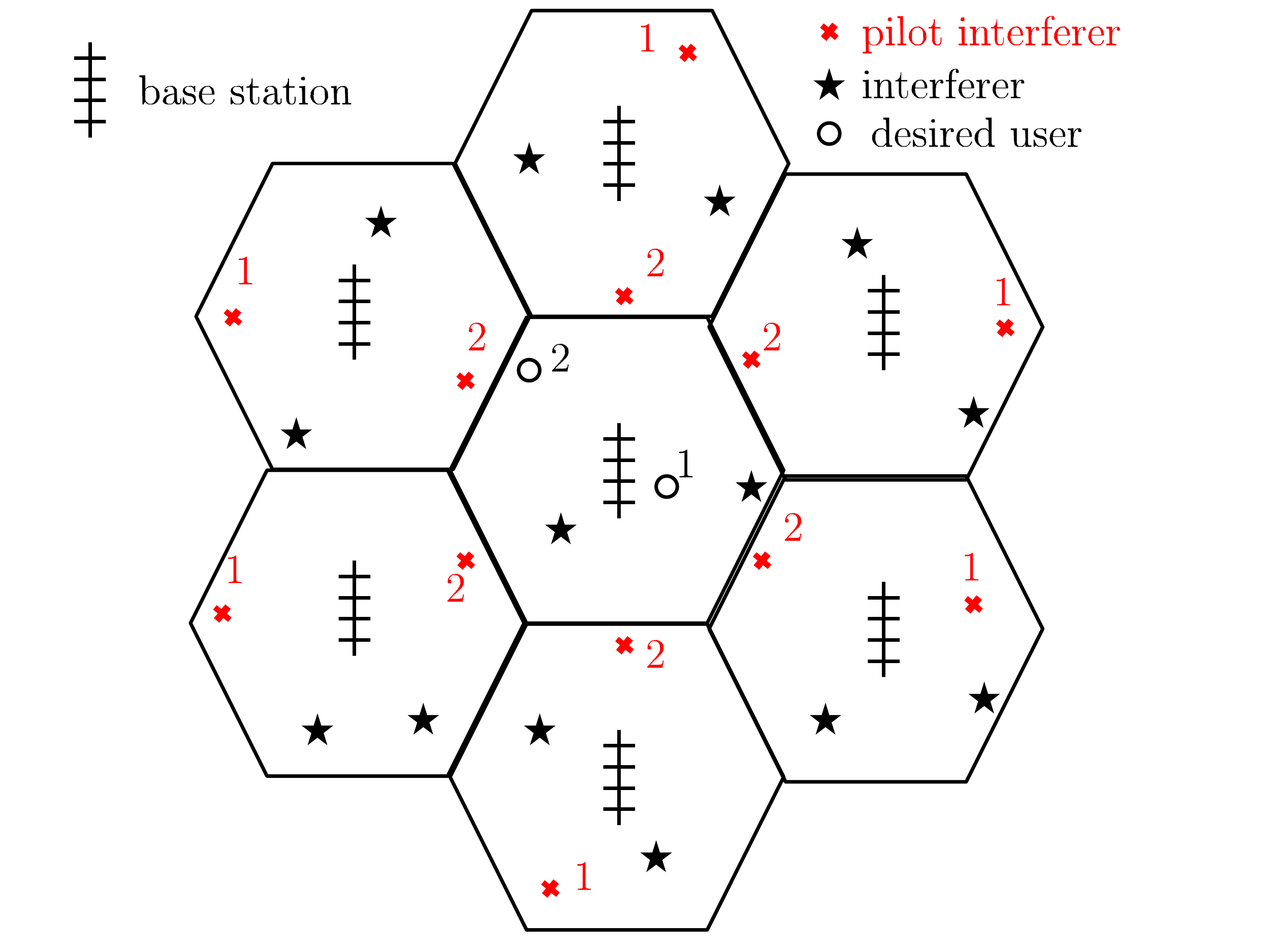}
      \caption{ In the favourable case, the sum of the received powers of interferers contributing to pilot contamination are very less as compared to that of the desired user. User $1$ in the center cell represents such a scenario. The SINR with a pilot corrupted estimate is then comparable to that of perfect estimate. On the other hand for user $2$ in the center cell, the pilot interferers received powers are comparable to that of the desired user and represents the worst case scenarios.} 
      \label{fig:7cellSys}
    \end{center}
\end{figure}

For the numerical evaluation, we consider hexagonal cells with users uniformly distributed in each of the cells as shown in Fig. \ref{fig:7cellSys}. We consider a set up where $6$ closest cells are interfering with the center cell. We assume received powers from all the users are unity i.e $\beta_{jk} = 1$. We consider the SINR for the user one in the center cell. Fig. \ref{fig:AsyPlots} plots the asymptotic SINR of the matched filter with a perfect estimate and that of a matched filter with a pilot corrupted estimate for the case of equal received powers. It is seen that using a pilot interference based estimate for matched filtering causes at least a $10$~dB loss in the SINR. This is dependent on the users contributing to pilot interference based estimate and hence dependent on $B$. Fig. \ref{fig:m50_TH} shows that the random realizations of the SINR for different values of $\alpha$ are clustered around the asymptotic plots.

%
%
\label{sec:results}

\begin{figure}[h]
    \begin{center}
      \includegraphics[width = 3in, clip = true, trim = 100 240 100 240]{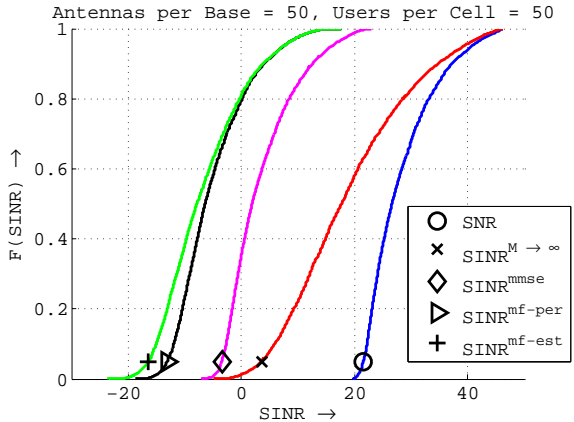}
      \caption{ The figure shows the CDF of SINR for five different system situations is used to compare the effect of interference when the number of antennas and users are comparable. We can see that MMSE performs much better than matched filter with a corrupt estimate. } 
      \label{fig:CDFSINR}
    \end{center}
\end{figure}

In order to get more intuition under practical scenarios of large scale fading gains, we consider the seven cell model with cell radius is $R = 2$~km, and assume a COST231 model for propagation loss between the base station and the users. The noise power is assumed to be $-174$~dBm and user transmit power of $23$~dBm. We plot the CDF of the SINR in Fig. \ref{fig:CDFSINR} for the following scenarios, 
\begin{itemize}
\item $\alpha =0$ with $\kappa = 0$ corresponding to matched filtering with perfect estimate with number of antennas far exceeding the number of user.
\item $\alpha = 0$ and $\kappa = 1$ corresponding to case when we employ matched filtering at the receiver with a pilot corrupted channel estimate similar to  \cite{Marzetta2011},
\item $\alpha = 1$ with $50$ antennas and $50$ users per base station and $\kappa =0$, representing the case of matched filter receiver with a perfect channel estimate,
\item $\alpha = 1$ with $50$ antennas and $50$ users per base station and $\kappa =1$ as representing the case of matched filtering with a pilot corrupted channel estimate but with comparable number of antennas and users.
\end{itemize}
Again, we only consider evaluating the SINR of the first user at the first base station as the SINR of all other users in the system follow the same statistics. As compared to the case when $\alpha = 0$ there is at least $25$~dB loss in SINR when the number of antenna becomes equal to number of users. In the case of matched filtering with perfect estimate this can be explained entirely by the non-zero interference power because $\alpha =1$. When matched filtering with a pilot corrupted channel estimate is employed, the shift is due to the combination of both pilot interference power and that interference due to $\alpha =1$. More specifically consider a scenario when
\begin{eqnarray}
\label{eq:favScen}
\frac{\sum^{B}_{j=2} \beta_{j1}}{\beta_{11}} \ll 1 \Rightarrow  \frac{\sum^{B}_{j=2} \beta^2_{j1}}{\beta_{11}} \ll1
\end{eqnarray}
This corresponds to the fact that sum of received powers of the interferers are much less that that of desired user power. Under these conditions the SINR of the received signal in equation (\ref{eq:SINRLargeSystem}) is,
\begin{eqnarray}
\mathsf{SINR}^* &\approx&    \frac{ \rho_r\beta_{11} }{ 1 +   \rho_r    B\alpha \mathbb{E}[\beta]} \nonumber
\end{eqnarray}
and hence the SINR of the matched filter with the corrupt channel estimate is as good as the SINR with a perfect channel estimate. This is seen in the best case SINR realizations in Fig. \ref{fig:CDFSINR}. In Fig. \ref{eq:SINRMF}, the situation of user $1$ in the center cell represents the favourable scenario with the interferers contributing to the pilot contamination are far such that the condition (\ref{eq:favScen}) is satisfied. 

On the other hand if all the interferers gains are comparable to that of the desired users gains represented by
\begin{eqnarray}
\frac{\sum^{B}_{j=2} \beta_{j1}}{\beta_{11}} \approx B - 1
\end{eqnarray}
then pilot interference contributes negatively to the SINR in addition to interference averaging. This is seen in Fig. \ref{fig:CDFSINR} for the worst case SINR realizations. In Fig. \ref{fig:7cellSys}, the situation of desired user $2$ represents such a scenario with desired user $2$ at the edge of the center cell and the interferers contributing in the pilot estimate as shown in the figure. These observations are true for all values of $\alpha$. 

\section{Adaptive MMSE Filtering}
We propose that adaptive MMSE filtering can be used for each user to decode its uplink data. Adaptive MMSE does not require the explicit knowledge of the channel. Instead, pilot training sequences known at the base station are sent by every user simultaneously so that each user converges to its desired MMSE filter directly. The explanation for the implementation of adaptive MMSE is as follows. Given the received signal in equation (\ref{eq:RcvdSgnlVectorForm}), we wish to design an MMSE filter  $\mathbf{c}_{11}$ for the first user in the first cell so that $\mathbb{E}[\left| q_{11} - \mathbf{c}^{H}_{11}\mathbf{y}^{} \right|^{2}]$ is minimized. From \cite{VerduMUD}, the filter which minimizes mean square objective is given by $\mathbf{c}_{11} = (\mathbb{E}[\mathbf{y}^{}\mathbf{y}^H])^{-1} \mathbb{E}[q^{*}_{11}\mathbf{y}]$. In order to form the filter $\mathbf{c}_{11}$ directly, the channel knowledge $\mathbf{g}_{jk}$ from the first base to all the users is required.
A way to overcome this difficulty is by using stochastic gradient descent algorithm, so that the filter adaptively converges to MMSE as proposed for CDMA systems \cite{VerduMUD}. With $0 \leq t \leq T$ being the index of time-frequency resources spent for training data,  $\{\psi_{11}[t]\}$ the $T$ length training sequence for user indexed one and $\mu >0$, the progressively decreasing step size, a step in the algorithm is given by the equation,
\begin{eqnarray}
\label{eq:MMSEiteration}
\mathbf{c}_{11}[t] = \mathbf{c}_{11}[t-1] - \mu (\psi_{11}[t] - \mathbf{c}^H_{11}[t-1]\mathbf{y}^{}[t])\mathbf{y}^{}[t] \text{.}
\end{eqnarray}
It has been shown that there exist positive values of $\mu$ so that the iteration (\ref{eq:MMSEiteration}) converges to the MMSE filter. Also, as long as training data satisfies $\mathbb{E}[\psi_{jk}\psi^*_{li}]= 0$ for $\{j,k\} \neq \{l,i\}$ the filter converges to the MMSE of user $1$ in cell $1$. 
Consequently, the SINR of the MMSE estimate is given by,
\begin{eqnarray}
\mathsf{SINR}^{\mathrm{mse}}_{} = \frac{\rho_r}{M} {\mathbf{g}}^{H}_{11}\left( \frac{\rho_r}{M} \underset{\{(j,k) \neq(1,1) \}}{\sum \sum} {\mathbf{g}}^{}_{jk}  {\mathbf{g}}^{H}_{jk}  + I \right)^{-1} {\mathbf{g}}^{}_{11}
\end{eqnarray}

In Fig.~\ref{fig:CDFSINR} we plot the SINR of the MMSE filter assuming that the algorithm has converged to $\mathbf{c}_{11}$. As compared to the SINR with matched filtering, when $\alpha =1 $, it is seen that MMSE can provide a gain of at least $10$~dB because of interference cancelling capabilities.

\section{Conclusion}
In this work we consider a multi-cell multi-antenna system with large number of antennas and users.  We perform a large system analysis and derive the asymptotic SINR (equation (\ref{eq:SINRLargeSystem})) when a matched  filter with a pilot contaminated estimate is employed at the receiver. We verified this expression numerically by seeing that the random realizations of SINR converge to the asymptotic limit as we increase the number of antenna. We also showed that when the number of users is comparable to the number of antennas the performance of the matched filter with corrupt estimate is limited by the interference. This is in contrast with the case when number of antennas far exceed the number of users, where the pilot interference term has the effect of only reducing the $5$ percentile SINR. We also proposed an adaptive MMSE symbol detection scheme where there is no necessity to estimate the channel. This could help in mitigating the effect of pilot contamination in the uplink. However, since the training symbols are limited due to coherence time of the channel, convergence to the MMSE filter will be an issue and this is a subject of current research. 

\appendices

\section{Proof of Theorem \ref{th:SINRLargeSystem}}

In equation (\ref{eq:SINRMF}) numerator term can be simplified as,
\begin{align}
 \frac{\hat{\mathbf{g}}^H_{11}\mathbf{g}_{11}}{M} &=    \frac{( \mathbf{g}^H_{11} + \sqrt{\kappa} \sum^B_{j=2}\mathbf{g}_{j1})^H \mathbf{g}_{11} }{M} \\
&= \beta^2_{11}\frac{\mathbf{h}^H_{11} \mathbf{h}_{11}}{M} + \sqrt{\kappa} \sum^B_{j=2}\sqrt{\beta_{j1}\beta_{11}}\frac{\mathbf{h}^H_{j1}\mathbf{h}_{11}}{M} \label{eq:SINRMFNum}
\end{align}
From (\ref{eq:SINRMFNum}), we observe that, $\lim_{M \rightarrow \infty} \hat{\mathbf{g}}^H_{11}\mathbf{g}_{11}/M = \beta_{11}$.
In equation (\ref{eq:SINRMF}), the first term in the denominator which is the noise term given by $\hat{\mathbf{g}}^H_{11} \hat{\mathbf{g}}_{11}/M$ is equal to,
\begin{align}
&\frac{1}{M}\left( \mathbf{g}_{11} +\sqrt{\kappa}\sum^B_{j=2}\mathbf{g}_{j1} \right)^H \left( \mathbf{g}_{11} + \sqrt{\kappa}\sum^B_{s=2}\mathbf{g}_{s1}\right)  \\
&= \frac{ \mathbf{g}^H_{11}\mathbf{g}_{11}}{M} + \kappa \sum^B_{j=2}\sum^B_{s = 2}\frac{\mathbf{g}^H_{j1}\mathbf{g}_{s1}}{M} + \sqrt{\kappa} \sum^B_{s=2} \frac{ \mathbf{g}^H_{11}\mathbf{g}_{s1}   }{M}  \nonumber \\
& \qquad+ \sqrt{\kappa} \sum^B_{j=2}\frac{\mathbf{g}^H_{j1} \mathbf{g}_{11}}{M}\label{eq:SINRMFDenoNoise}
\end{align}
From (\ref{eq:SINRMFDenoNoise}), we observe that, $\lim_{M \rightarrow \infty} \hat{\mathbf{g}}^H_{11}\mathbf{\hat{g}}_{11}/M = \beta_{11}+ \kappa \sum^B_{j=1}\beta_{j1}$. Converting the subscript of the channel vectors which is double indexed into subscript which is single indexed by using the definition $\mathbf{g}_{jk} = \sqrt{\beta_{jk}}\mathbf{h}_{jk} \triangleq \sqrt{\beta_{jk}}\mathbf{h}_{(j-1)K+k} = \mathbf{g}_{(j-1)K+k}$, the interference term in equation (\ref{eq:SINRMF}) 
can be simplified as,
\begin{align}
&\frac{\rho_r}{M^2} \underset{(j,k)\neq(1,1)}{\sum_j\sum_k}  \left| \left( \mathbf{g}_{11} + \sqrt{\kappa}\sum^B_{i=2}\mathbf{g}_{i1} \right)^H \mathbf{g}_{jk}  \right|^2 \nonumber \\
&= \frac{ \rho_r}{M^2} \sum^{BK}_{i=2} \left|  \left(  \mathbf{g}_1+ \sqrt{\kappa}\sum^{B-1}_{u=1} \mathbf{g}_{uK+1} \right)^H \mathbf{g}_i \right|^2 \nonumber \\
&= \frac{\rho_r }{M^2}  \sum^{BK}_{i=2}  \left| \mathbf{g}^H_1 \mathbf{g}_i+ \sqrt{\kappa}\sum^{B-1}_{u=1}\mathbf{g}^H_{uK+1} \mathbf{g}_i \right|^2 \nonumber \\
&= \frac{\rho_r }{M^2}  \sum^{BK}_{i=2}  \left( \left| \mathbf{g}_1\mathbf{g}_i \right|^2 + \sqrt{\kappa}\sum^{B-1}_{s=1}\mathbf{g}^H_1\mathbf{g}_i\mathbf{g}^T_{sK+1} \mathbf{g}^*_i  \right. \nonumber \\
& \qquad + \sqrt{\kappa}\sum^{B-1}_{u=1}\mathbf{g}^H_{uK+1} \mathbf{g}_i\mathbf{g}^T_1\mathbf{g}^*_i \left.+ \kappa \sum^{B-1}_{u=1} \sum^{B-1}_{s=1} \mathbf{g}^H_{uK+1} \mathbf{g}_i \mathbf{g}^T_{sK+1} \mathbf{g}^{*}_i \right)\nonumber \\
&= \frac{\rho_r }{M^2}  \sum^{BK}_{i=2}  \left( \left| \mathbf{g}_1\mathbf{g}_i \right|^2 + \kappa \sum^{B-1}_{u=1} \left |\mathbf{g}^H_{uK+1} \mathbf{g}_i \right|^2    \right. \nonumber \\
& \qquad + \sqrt{\kappa}\sum^{B-1}_{s=1}\mathbf{g}^H_1\mathbf{g}_i\mathbf{g}^T_{sK+1} \mathbf{g}^*_i +\sqrt{\kappa}\sum^{B-1}_{u=1}\mathbf{g}^H_{uK+1}\mathbf{g}_i\mathbf{g}^T_1\mathbf{g}^*_i  \nonumber \\
& \qquad \left.+ \kappa \underset{u \neq s}{\sum^{B-1}_{u=1} \sum^{B-1}_{s=1}} \mathbf{g}^H_{uK+1} \mathbf{g}_i \mathbf{g}^T_{sK+1} \mathbf{g}^{*}_i \right) \nonumber
\end{align}

Using $\overset{p}{\longrightarrow}$ to denote convergence in probability, from \cite[Proposition~3.3]{TseHanly1999} we can show that in the limit of $M \rightarrow \infty$, $K \rightarrow \infty$, with $K/M = \alpha$,
\begin{align}
\frac{\rho_r }{M^2}  \sum^{BK}_{i=2}   \left| \mathbf{g}^H_1\mathbf{g}_i \right|^2 &\overset{p}{\longrightarrow} \rho_r \beta_{11} B \alpha \mathbb{E}[\beta].
\end{align}
Also, for $u \in \{1,2 \dots B-1 \}$
\begin{align}
&\frac{\rho_r }{M^2} \sum^{BK}_{i=2} \left| \mathbf{g}^H_{uK+1}\mathbf{g}_i \right|^2 ,\nonumber \\
&=  \frac{\rho_r }{M^2}  \sum^{}_{i\neq 1,uK+1} \left| \mathbf{g}^H_{uK+1}\mathbf{g}_i \right|^2 + \frac{\rho_r}{M^2}\left||\mathbf{g}_{uK+1} \right||^4, \label{eq:firstTermTH} \\
&\overset{p}{\longrightarrow} \rho_r \beta_{u+1,1} B \alpha \mathbb{E}[\beta] + \rho_r \beta^2_{u+1,1}, \nonumber
\end{align}
where, in equation (\ref{eq:firstTermTH}) the first term converges due to \cite[Proposition~3.3]{TseHanly1999} and the second term converges due to weak law of large numbers. Assuming $a \neq b$, the other terms are of the form,
\begin{align}
&\frac{\rho_r }{M^2}  \sum^{BK}_{i=2} \mathbf{g}^H_{a}\mathbf{g}_i \mathbf{g}^T_{b} \mathbf{g}^{*}_i  \nonumber \\
& = \frac{\rho_r }{M^2}  \sum^{}_{i \neq 1,a,b} \mathbf{g}^H_{a}\mathbf{g}_i \mathbf{g}^T_{b} \mathbf{g}^{*}_i  \nonumber \\
& \qquad + \frac{\rho_r}{M^2} || \mathbf{g}_a||^2 \mathbf{g}^T_b\mathbf{g}^*_a + \frac{\rho_r}{M^2} || \mathbf{g}_b||^2 \mathbf{g}^H_a\mathbf{g}_b \label{eq:zerotermTH}
\end{align}
In a similar approach to analysis of the interference term in \cite[Proposition~3.3]{TseHanly1999} we can show that in equation (\ref{eq:zerotermTH}) the first term converges to its expectation which is $0$ and the second and third term converges to $0$ by weak law of large numbers. Since, each of the above terms converge in probability, the sum of the terms also converge in probability. After rearranging the terms we get the expression of SINR as in Theorem \ref{th:SINRLargeSystem}.

\bibliographystyle{IEEEbib}

\end{document}